\documentclass[a4paper,11pt]{article}

\usepackage{jheppub} 

\usepackage[T1]{fontenc} 
\usepackage{slashed}
\usepackage{diagbox}
\usepackage{makecell}

\title{\boldmath Probing compressed higgsinos with forward protons at the LHC}

\author[a]{Hang Zhou,}
\author[b]{Ning Liu}

\affiliation[a]{School of Microelectronics and Control Engineering, Changzhou University\\Changzhou, China}
\affiliation[b]{Physics Department and Institute of Theoretical Physics, Nanjing Normal University\\Nanjing, China}

\emailAdd{zhouhang@cczu.edu.cn}
\emailAdd{liuning@njnu.edu.cn}

\abstract{Supersymmetric models with nearly-degenerate light higgsinos provide a consistent solution to the naturalness problem under rigorous constraints from current experimental searches for sparticles. However, it is challenging to probe the compressed scenarios at collider experiments due to the hard-to-detect soft final states. To overcome this issue, strategies have been proposed to take advantage of the photon fusion along the ultraperipheral collision at the Large Hadron Collider, which are feasible with the forward detectors installed at the ATLAS and CMS detectors. In this paper, we demonstrate a search strategy for the chargino pair production via photon fusion at the 13 TeV LHC, realizing a good sensitivity that can exclude $m_{\tilde{\chi}^{\pm}_{1}}$ up to 270 GeV (308 GeV) for a mass splitting $\Delta m(\tilde{\chi}^{\pm}_{1},\tilde{\chi}^{0}_{1})\in[1,15]$ GeV with an integrated luminosity of 100 fb$^{-1}$ (3 ab$^{-1}$) at 95\% C.L.}

\begin{document}
\maketitle
\flushbottom

\section{Introduction}
\label{sec:intro}

Supersymmetric (SUSY) extensions beyond the Standard Model (SM) not only provide a solution to alleviate the naturalness problem, but the lightest supersymmetric particle (LSP) can also constitute the dark matter in the universe. Therefore, SUSY models have been widely studied and searched for in various experiments, especially the high-energy collider experiments. However, current results from the Large Hadron Collider (LHC) have put stringent lower constraints on the masses of the SUSY particles and excluded quite a large parameter space for low-energy SUSY. With the general assumptions of R-parity conservation and the neutralino $\tilde{\chi}^{0}_{1}$ being the LSP, the scalar top quark (stop) with its mass larger than 1.25 TeV is excluded at 95\%C.L. in the context of simplified SUSY models \cite{ATLASCollaboration2020a}, while the lower mass bound for the gluino has been put at 2 TeV \cite{ATLASCollaboration2020}. As for electroweakinos, the search for pair production of charginos or neutralinos using fully hadronic final states \cite{ATLAS:2021yqv} has excluded the wino\,(higgsino) mass below 1060\,(900) GeV assuming the LSP mass below 400\,(240) GeV and the mass splitting larger than 400\,(450) GeV. For the mass splitting near the electroweak scale, the search for chargino-neutralino pair production has excluded the mass range below 345 GeV \cite{Aad2020}. With a moderate mass splitting ($\sim$ a few GeV) between chargino and neutralino, the lower limit on chargino mass varies from 193 to 240 GeV depending on different simplified models \cite{Aad2020a}, while for a more compressed spectrum (1.5 to 2.4 GeV), the LEP2 bound on chargino mass at 92.4 GeV remains the strongest \cite{LEPbound}.

It follows then that the compressed scenarios have been subject to a relatively loose restriction from the collider experiments, which are also well-motivated taking into account the requirement of naturalness in the Minimal Supersymmetric Standard Model (MSSM). Consider the minimization of the Higgs potential at tree-level related to the mass of $Z$ boson \cite{Arnowitt1992}
\begin{align}
\frac{m^{2}_{Z}}{2}=\frac{m^{2}_{H_{d}}-m^{2}_{H_{u}}\tan^{2}\beta}{\tan^{2}\beta-1}-\mu^{2},
\label{minimal_potential}
\end{align}
where $\tan\beta=v_{u}/v_{d}$ is defined as the ratio between the VEVs of two Higgs doublets $H_{u}$ and $H_{d}$, the soft SUSY breaking masses of which at weak scale are denoted by $m^{2}_{H_{d}}$ and $m^{2}_{H_{u}}$. $\mu$ refers to the mass parameter of higgsino. To arrive at the correct Z mass without resort to large fine-tuning, terms in Eq.~\eqref{minimal_potential} are expected to be in a comparable magnitude \cite{Barbieri1988} and moderate values of $m_{H_{u}}$ and $\mu$ around a few hundreds GeV are preferred, hence leading to light higgsinos in the mass spectrum which have been studied phenomenologically in various ways \cite{Han2014,Hikasa2016,Liu2017,Han2018,Harland-Lang2019,Han2020,Godunov2020,Lv2022,Abdughani2018,Abdughani2019,Abdughani2020}. Whereas, given the above-mentioned exclusion limits of $1\sim2$ TeV on gluino and stop masses as well as the observed 125 GeV Higgs boson \cite{Aad2012,Chatrchyan2012}, other scenarios that can reconcile the observations with naturalness requirement are drawing more attention, including the radiative natural SUSY \cite{Baer2012} and minimal supergravity in the focus point or hyperbolic branch regions \cite{Chan1998,Feng2000}. Such scenarios are generally known as higgsino world \cite{Kane1998} in which a low fine-tuning can be realized with heavy scalars of a few TeV and a low value of $\mu$ around sub-TeV. In the limit of $\mu$ being much less than the gauginos masses $M_{1,2}$ for bino and wino, respectively, which are favored to be in the same magnitude of the gluino mass $M_{3}$ \cite{Giudice1996,Martin2009,Akula2013,Gogoladze2013}, the difference between the lightest chargino and neutralino can be expressed as \cite{Giudice1996}
\begin{align}
\Delta m = m_{\tilde{\chi}^{\pm}_{1}} - m_{\tilde{\chi}^{0}_{1}}=\frac{m^{2}_{W}}{2M_{2}}\tan^{2}\theta_{W}(1+\sin2\beta)+\frac{m^{2}_{W}}{2M_{1}}\left(1-\sin2\beta-\frac{2\mu}{M_{2}}\right),
\end{align}
where $m_{W}$ is the $W$ boson mass and $\theta_{W}$ the Weinberg angle, leading to nearly degenerate and light electroweakinos with a mass splitting being a few GeV (assuming a large $\tan\beta\gtrsim10$).

Compared to the colored sparticles, charginos and neutralinos have a relatively low discovery potential at the hadron colliders as they do not participate in the strong interaction and thus have smaller production cross sections through electroweak interaction. Furthermore, it is particularly difficult to search for signals of the compressed mass spectrum through pair production of the electroweakinos, the challenges of which mainly come from the similar topology of $WW$ events as the SM background, as well as the soft final states with low momenta commonly below the threshold of detector acceptance. Additional jets or photons from the initial state radiation (ISR) are generally required to trigger the signal events, the effect of which, however, is constrained by the large systematic uncertainties at the high-luminosity LHC (HL-LHC) \cite{Han2014,Schwaller:2013baa,HanZ2014,Baer2014,Aad2019,Barducci2015}. For an even more squeezed mass spectrum around hundreds of MeV, searches have been proposed for signals from disappearing tracks \cite{Mahbubani:2017gjh,Fukuda:2017jmk}.

In consideration of these facts, search strategies using the LHC as a photon-photon collider have been studied recently as a promising way to search for the light and nearly degenerate spectrum of electroweakinos and sleptons \cite{Beresford2019,Harland-Lang2019,Godunov2020}, which had been made possible with the development of the forward detectors installed at the LHC including the ATLAS Forward Proton \cite{Tasevsky2015,AFP2015} and CMS-TOTEM Precision Proton Spectrometer \cite{CTPPS2006,CTPPS2014}, known as the AFP and CT-PPS, respectively. These forward detectors are located near the colliding beam at the distance $\sim220$m from the collision point, aiming to identify the outgoing protons that go through the ultraperipheral collision (UPC), in which the initial protons brush against each other and remain intact. Alongside the UPC, electromagnetic fields surrounding the colliding protons are approximately equivalent to colliding beams of on-shell photons, commonly known as the equivalent photon approximation \cite{Budnev1975}, leading to processes of central exclusive production (CEP)
\begin{align}
pp\to p+X+p\,,
\end{align}
where $X$ refers to a pair of electrically charged states produced through equivalent photon fusion. In the case of our study on searching for electroweakinos, the lightest charginos $\tilde{\chi}^{\pm}_{1}$ can be produced in pair $\gamma\gamma\to\tilde{\chi}^{+}_{1}\tilde{\chi}^{-}_{1}$, and then decay into neutralinos and $W$ bosons followed by leptons or jets in the final states. With the above-mentioned forward detectors, tagging of outgoing protons in such processes can help reconstruct the initial states and further determine the final missing momentum, realizing a better reconstruction of neutralino mass against the SM neutrino background. In this paper, by emulating the proton tagging as a way to suppress the SM background, we perform a study on the prospects of searching for electroweakinos under compressed scenarios via photon fusion at the 13 TeV LHC. In the next section, we analyze the signal and relevant SM background, as well as numerical simulation of both processes. In Section III, we present our search strategies and the results on the observability. Section IV is our conclusion.

\section{Signal and Simulation}
\begin{figure}[t]
\centering
\begin{minipage}{0.48\linewidth}
  \centerline{\includegraphics[scale=0.22]{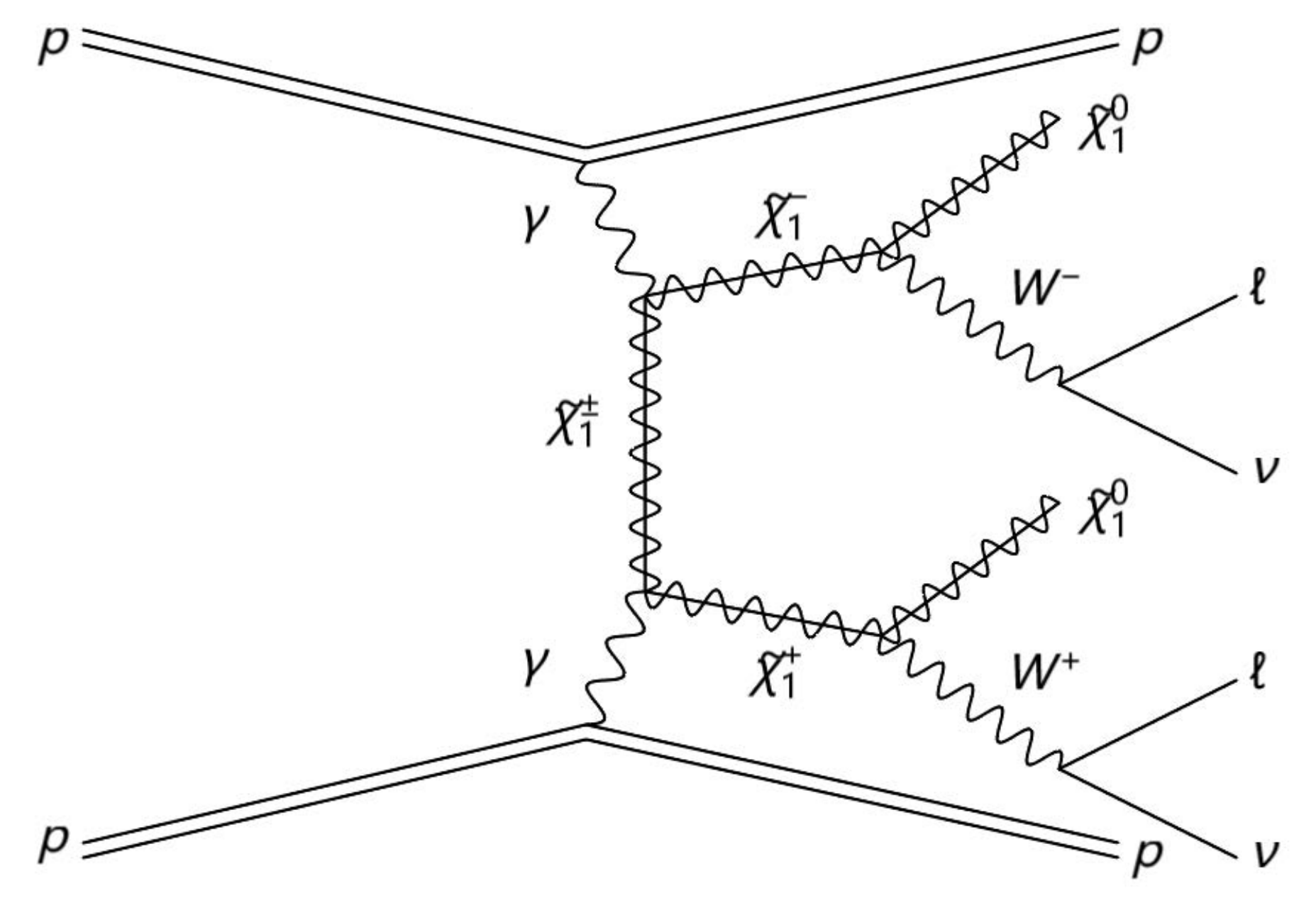}}
  \centerline{(a)}
\end{minipage}
\hfill
\begin{minipage}{0.48\linewidth}
  \centerline{\includegraphics[scale=0.22]{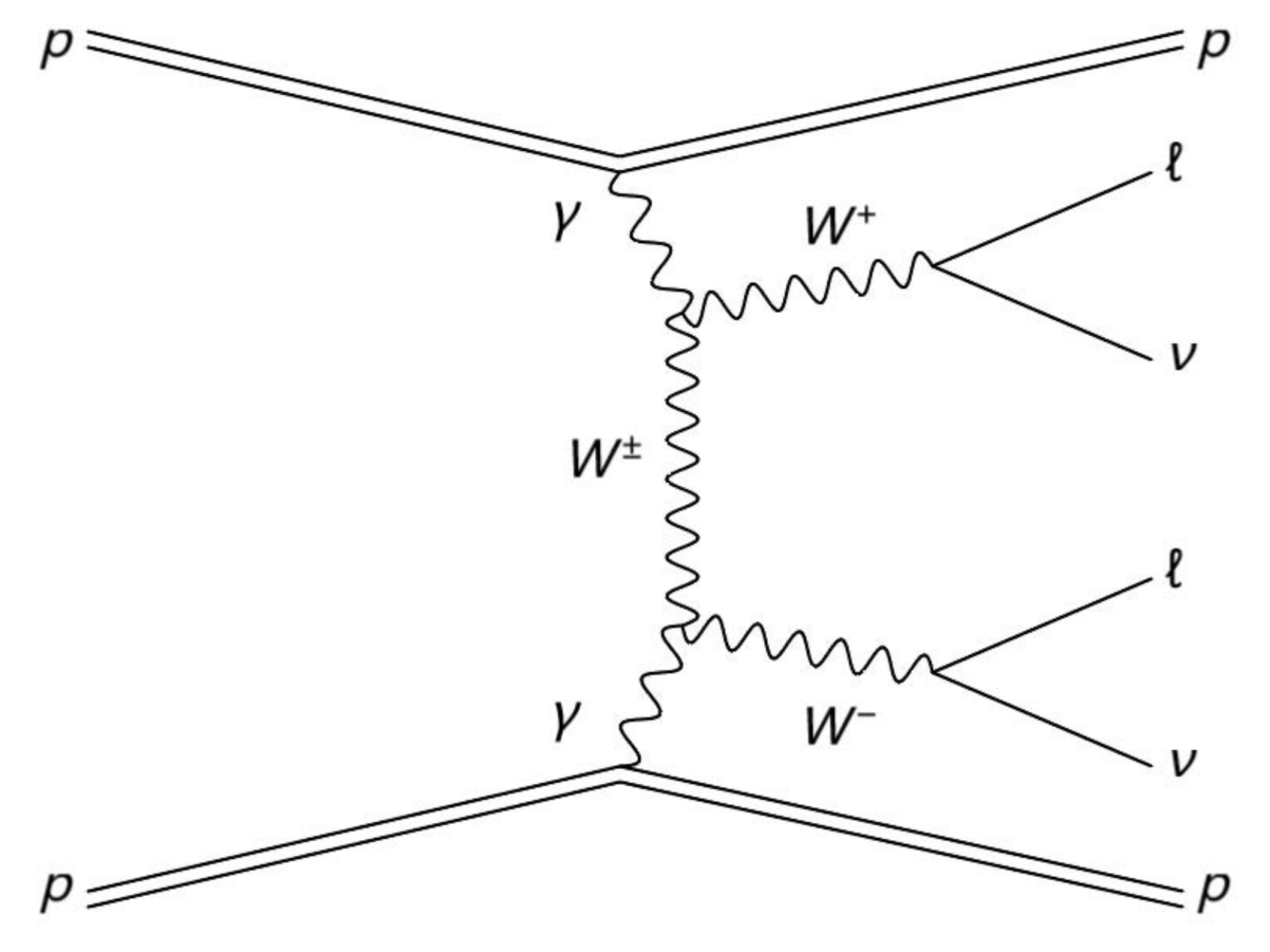}}
  \centerline{(b)}
\end{minipage}
\caption{Feynman diagrams of signal and the SM background processes.}
\label{fig:feyn}
\end{figure}

Photon fusions can achieve sufficient rates at the LHC to search for SUSY particles \cite{Ohnemus1994,Pierzchala2006,Schul2008,Harland-Lang2012,Harland-Lang2019,Beresford2019,Godunov2020} and other new physics \cite{Chapon2010,Fichet2014,Fichet2017,Knapen2017,Khoze2017,Baldenegro2018}. In the present study on pair production of charginos $\tilde{\chi}^{\pm}_{1}$ within the MSSM, the cross section via photon fusion can be expressed as following \cite{photonfusionXS}
\begin{align}
\sigma_{\gamma\gamma\to\tilde{\chi}^{+}_{1}\tilde{\chi}^{-}_{1}}=\frac{4\pi\alpha^{2}}{s}\left[\left(1+\lambda-\frac{\lambda^{2}}{2}\right)\ln\frac{1+\sqrt{1-\lambda}}{1-\sqrt{1-\lambda}}-(1+\lambda)\sqrt{1-\lambda}\right],
\end{align}
in which $\lambda=4m^{2}_{\tilde{\chi}^{\pm}_{1}}/s$ and $\sqrt{s}$ is the invariant mass of the chargino pair. For the chargino pair production along with the UPC, the convolution with the equivalent photon spectrum should be performed to arrive at the total cross section \cite{Godunov2020}
\begin{align}
\label{convol}
\sigma_{pp\to p(\gamma\gamma\to\tilde{\chi}^{+}_{1}\tilde{\chi}^{-}_{1})p}=\int\int\sigma_{\gamma\gamma\to\tilde{\chi}^{+}_{1}\tilde{\chi}^{-}_{1}}n_{p}(E_{1})n_{p}(E_{2})dE_{1}dE_{2},
\end{align}
where $E_{1,2}$ refer to the energies of the equivalent photons around the colliding protons and $n_{p}(E_{1,2})$ their spectra distributions. Note that integral limits should not be taken as $(0,+\infty)$ since the outgoing protons are detected by the forward detectors with certain acceptances dependant on their energy loss, hence the cross section should be calculated taking into account the equivalent photons with energy-dependant probabilities which we will discuss below. As our signal of the electroweakino pair production, we decay the charginos into neutralinos with the SM $W$ bosons, and, to realize an effective differentiation from the hadronic background at a hadron collider, specify the leptonic channels of $W$ decay $W^{+}W^{-}\to\ell^{+}\nu_{\ell}\ell^{-}\bar{\nu}_{\ell}$, $\ell=e, \mu$. The signal process is then
\begin{align}
\label{signal}
pp\to p\left(\gamma\gamma\to\tilde{\chi}^{+}_{1}\tilde{\chi}^{-}_{1}\to\tilde{\chi}^{0}_{1}\tilde{\chi}^{0}_{1}W^{+}W^{-}\to\tilde{\chi}^{0}_{1}\tilde{\chi}^{0}_{1}\ell^{+}\nu_{\ell}\ell^{-}\bar{\nu}_{\ell}\right)p\,,
\end{align}
the Feynman diagram of which is shown in \figurename~\ref{fig:feyn} (a). For the SM background, we consider the $WW$ events through the same UPC process followed by the $W$ leptonic decay as shown in \figurename~\ref{fig:feyn} (b),
\begin{align}
\label{bkg}
pp\to p\left(\gamma\gamma\to W^{+}W^{-}\to\ell^{+}\nu_{\ell}\ell^{-}\bar{\nu}_{\ell}\right)p\,.
\end{align}
To perform Monte-Carlo simulations, we generate the above signal and background processes by \textsc{MadGraph5\_aMC@NLO}(version 3.0.0)  \cite{Alwall2014} with the NN23LO1 PDF \cite{Ball2013}. The parton-level events then go through parton showering and hadronization via \textsc{pythia-8.2} \cite{Sjostrand2006}, as well as detector simulation with tuned \textsc{delphes-3.4.1} \cite{DeFavereau2014}, the whole procedure of which is conducted within the framework of \textsc{checkmate2} (version 2.0.26) \cite{Dercks2017}. SUSY mass spectra used in the event generation are calculated by the package SUSYHIT \cite{Djouadi:2006bz}

\begin{table}
\centering
\begin{tabular}{|c|*{5}{c|}}
\hline
\,$E_{\gamma}$\,(GeV)\, & \,(0,100]\, & \,(100,120]\, & \,(120,150]\, & \,(150,400]\, & \,(400,$+\infty$)\, \\ \hline
Eff. & 0 & 50\% & 70\% & 90\% & 80\% \\ \hline
\hline
\end{tabular}
\caption{Effective acceptance for the equivalent photons based on the energy-dependant proton tagging efficiencies \cite{AFP2015,CTPPS2006,CTPPS2014}.}
\label{tab:acceptance}
\end{table}
As stated in the preceding section, the ultraperipheral collision at the LHC is featured with intact outgoing protons, which are identified by the forward detectors with certain acceptances related to the proton's energy loss $\xi\equiv1-E_{\text{out}}/E_{\text{in}}$, where $E_{\text{out}}$ refers to the measured energies of outgoing protons and $E_{\text{in}}$ those of the ingoing protons. For $\xi\in(0.015, 0.15)$, the proton acceptance approximates to 100\%~\cite{AFP2015,CTPPS2006,CTPPS2014}, corresponding to energies of colliding photons $E_{\gamma}\in(100, 1000)$ GeV in the case of $\sqrt{s}=13$ TeV collisions. As phenomenological studies indicate lower values of acceptances $\sim90\%$ for intact protons after emitting photons \cite{Khoze2002}, we adopt conservatively the proton acceptance as 0\% for photons with energy of $E_{\gamma}<100$ GeV, which increases with the photon energy and then decreases to and remains 80\%, as listed in \tablename~\ref{tab:acceptance}. Alongside extracting the kinematic information of the photons, we smear their four-momenta by the Gaussian function with a width of 5\% according to the AFP resolution of 5 GeV~\cite{AFP2015}. The extraction and smearing of photon momenta from the parton-level events and the simulation of photon acceptance are realized by using the PYLHE package~\cite{pylhe}.

\section{Search Strategy and Results}

A major difference between the signal Eq. \eqref{signal} and the background Eq. \eqref{bkg} comes from the invisible system. The missing energy in the background events consists of neutrinos from the $W$ leptonic decay, while for the signal it is also comprised of massive neutralinos and thus is expected to distribute in a higher energy region. And as the final states of signal and background both come from photon fusion alongside the UPC of protons, larger missing energy makes softer charged leptons in the signal events than those in the background, so the momenta of the final leptons should exhibit distinguishable distributions to separate the signal and background.

Since the products of photon fusion, charginos $\tilde{\chi}^{\pm}_{1}$, assume a symmetric decay topology $\tilde{\chi}^{+}_{1}\tilde{\chi}^{-}_{1}\to\tilde{\chi}^{0}_{1}\tilde{\chi}^{0}_{1}W^{+}W^{-}\to\tilde{\chi}^{0}_{1}\tilde{\chi}^{0}_{1}\ell^{+}\nu_{\ell}\ell^{-}\bar{\nu}_{\ell}$, we can then achieve a better separation between the signal and background by reconstructing a mass bound $m_{\text{DM}}^{\text{max}}$ on the invisible system, namely the Harland-Lang-Kom-Sakurai-Stirling variable, which was first introduced and defined in Ref \cite{Harland-Lang2012} as
\begin{align}
\left(m_{\text{DM}}^{\text{max}}\right)^{2}=(p_{\ell_{1}}\cdot p_{\ell_{2}})\times\left[c_{c}-\frac{c^{2}_{b}}{4c_{a}}\right]\,,
\label{mDMmax}
\end{align}
where $p_{\ell_{1,2}}$ are four-momenta of final charged leptons $\ell_{1,2}$ and $c_{a,b,c}$ are defined as
\begin{align}
c_{a}={}&\frac{1}{4}\frac{(\Lambda_{1}+\Lambda_{2})^{2}-2\Lambda_{\gamma\gamma}}{\Lambda_{\gamma\gamma}-2\Lambda_{1}\Lambda_{2}}\,,\\
c_{b}={}&\frac{1}{2}(\Lambda_{1}+\Lambda_{2}-2)\,,\\
c_{c}={}&\frac{1}{4}(\Lambda_{\gamma\gamma}-2\Lambda_{1}\Lambda_{2})\,,
\end{align}
with
\begin{align}
\Lambda_{1}=\frac{p_{\gamma\gamma}\cdot p_{\ell_{1}}}{p_{\ell_{1}}\cdot p_{\ell_{2}}}\,,\qquad\Lambda_{2}=\frac{p_{\gamma\gamma}\cdot p_{\ell_{2}}}{p_{\ell_{1}}\cdot p_{\ell_{2}}}\,,\qquad\Lambda_{\gamma\gamma}=\frac{m^{2}_{\gamma\gamma}}{p_{\ell_{1}}\cdot p_{\ell_{2}}}\,,
\end{align}
in which $p_{\gamma\gamma}$ is the sum of four-momenta of the two equivalent photons and $m_{\gamma\gamma}$ is the invariant mass of the diphoton system. The reconstruction of these kinematic variables relevant to the diphoton can be realized by the forward detectors measuring the outgoing protons, while in our simulations the information of the diphoton can be fetched directly from the MadGraph output. To emulate realistic detectors, diphoton momenta have been processed according to the proton tagging efficiencies and smeared as discussed at the end of the previous section. It should be noted that the present definition of $m_{\text{DM}}^{\text{max}}$ implies a slightly different mass bound from that defined in Ref~\cite{Harland-Lang2012}: the invisible system consists of neutralinos as well as neutrinos in our case, while the original definition in Ref~\cite{Harland-Lang2012} refers uniquely to the dark matter components. But the cutflow which we will show below suggests that this variable can also serve as an effective cut in our search.

\begin{table}
\centering
\begin{tabular}{|c|*{2}{c|}}
\hline
\diagbox{cuts}{SR} & $\Delta m=1\sim15$\,GeV & $\Delta m=20$\,GeV \\ \hline
cut-1 & $\slashed{E}_{T}\notin$\,[40,220]\,GeV & $\slashed{E}_{T}\notin$\,[40,290]\,GeV \\ \hline
cut-2 & $P_{T}(\ell_{1})\in$[2,5]\,GeV & $P_{T}(\ell_{1})\in$[2,10]\,GeV \\ \hline
cut-3 & \,\,$m_{\text{DM}}^{\text{max}}\in[110,260]$\,GeV\,\, & \,\,$m_{\text{DM}}^{\text{max}}\in[120,260]$\,GeV\,\, \\ \hline
\hline
\end{tabular}
\caption{Cuts applied to events with identification of initial two photons and at least two final opposite-sign leptons.}
\label{tab:cuts-1}
\end{table}

In consideration of the above analysis, we first select events that can be identified by the forward detectors, that is, we keep a certain number of events based on probabilities presented in \tablename~\ref{tab:acceptance} as a consequence of the energy-dependant proton acceptance rates. Besides the identification of two intact protons, we require two or more charged leptons in the final states against the hadronic background at the LHC and two of them should have opposite signs. These procedures are denoted by "Diphoton \& Dilepton" in our event selection listed in \tablename~\ref{tab:cuts-1} and can be regarded as a pre-selection followed by the cuts on three kinematic variables including the missing transverse energy $\slashed{E}_{T}$, transverse momentum of the leading lepton $P_{T}(\ell_{1})$ and the invisible system mass bound $m_{\text{DM}}^{\text{max}}$ defined in Eq. \eqref{mDMmax}.

\begin{figure}[t]
\centering
\begin{minipage}{0.48\linewidth}
  \centerline{\includegraphics[scale=0.44]{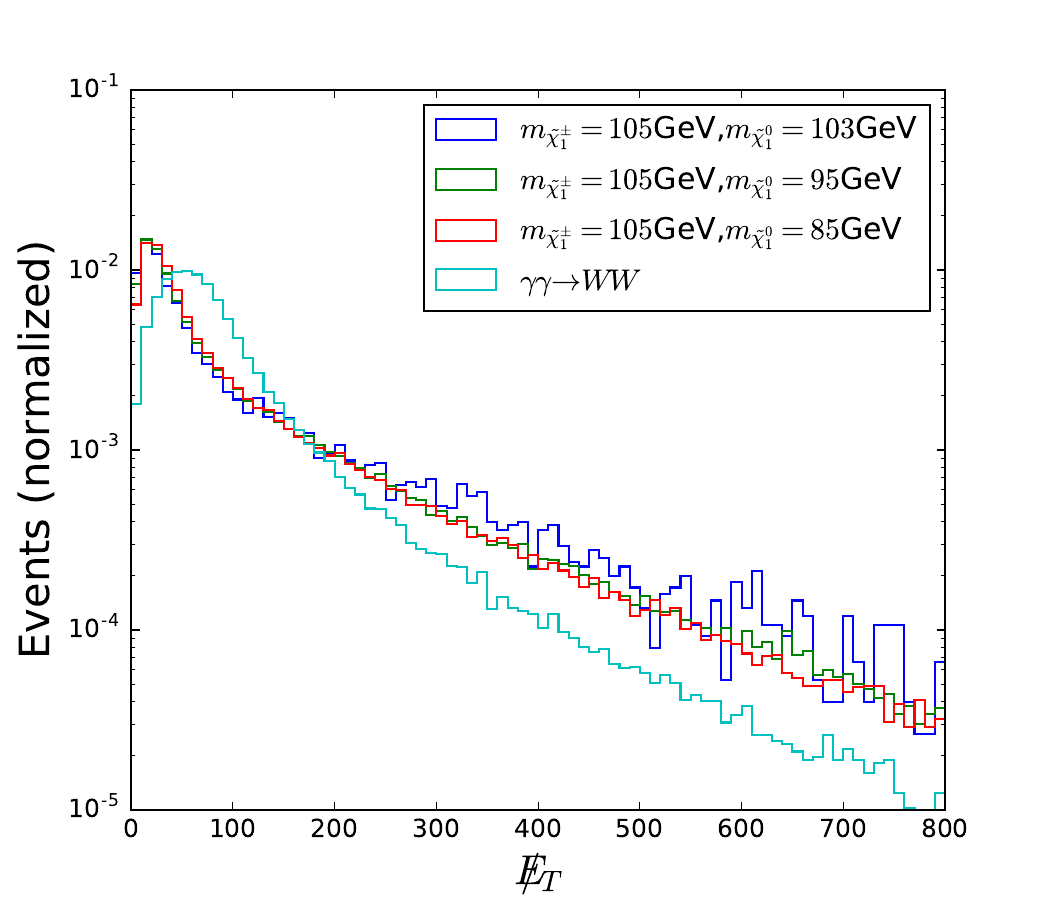}}
  \centerline{(a)}
\end{minipage}
\hfill
\begin{minipage}{0.48\linewidth}
  \centerline{\includegraphics[scale=0.44]{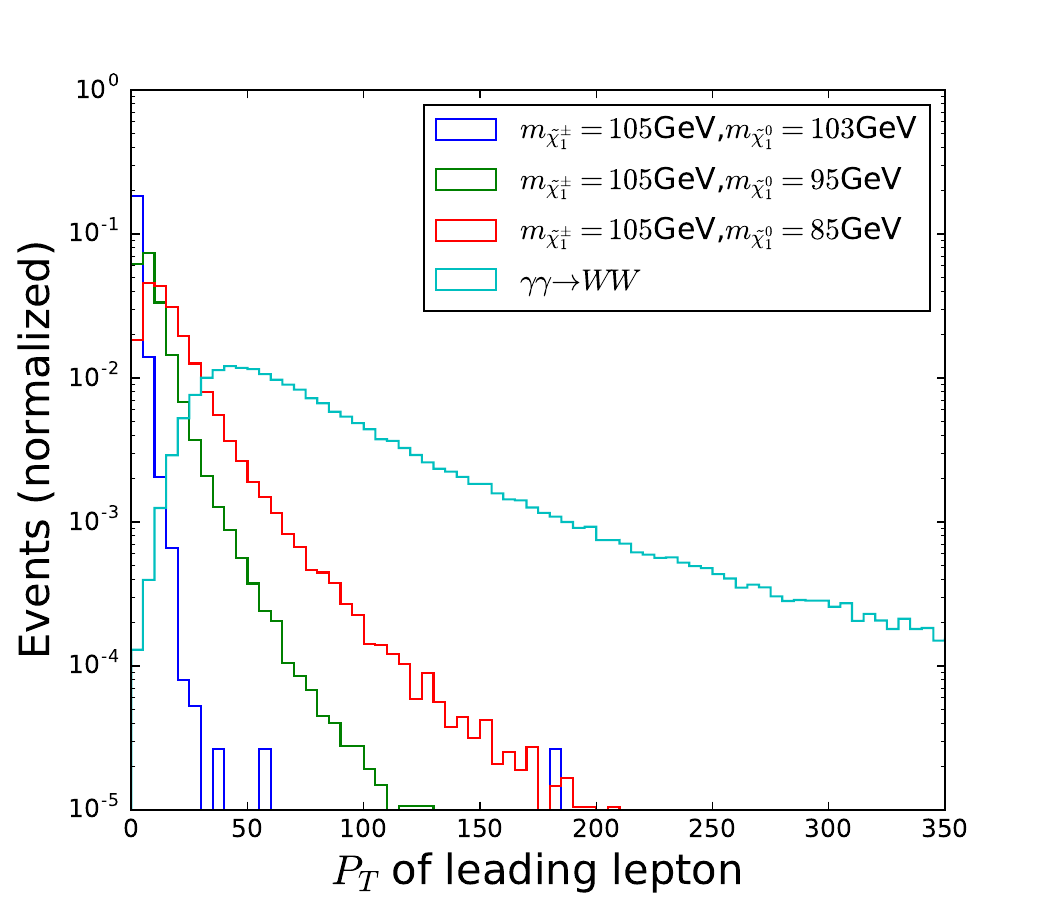}}
  \centerline{(b)}
\end{minipage}
\\[12pt]
\begin{minipage}{0.48\linewidth}
  \centerline{\includegraphics[scale=0.47]{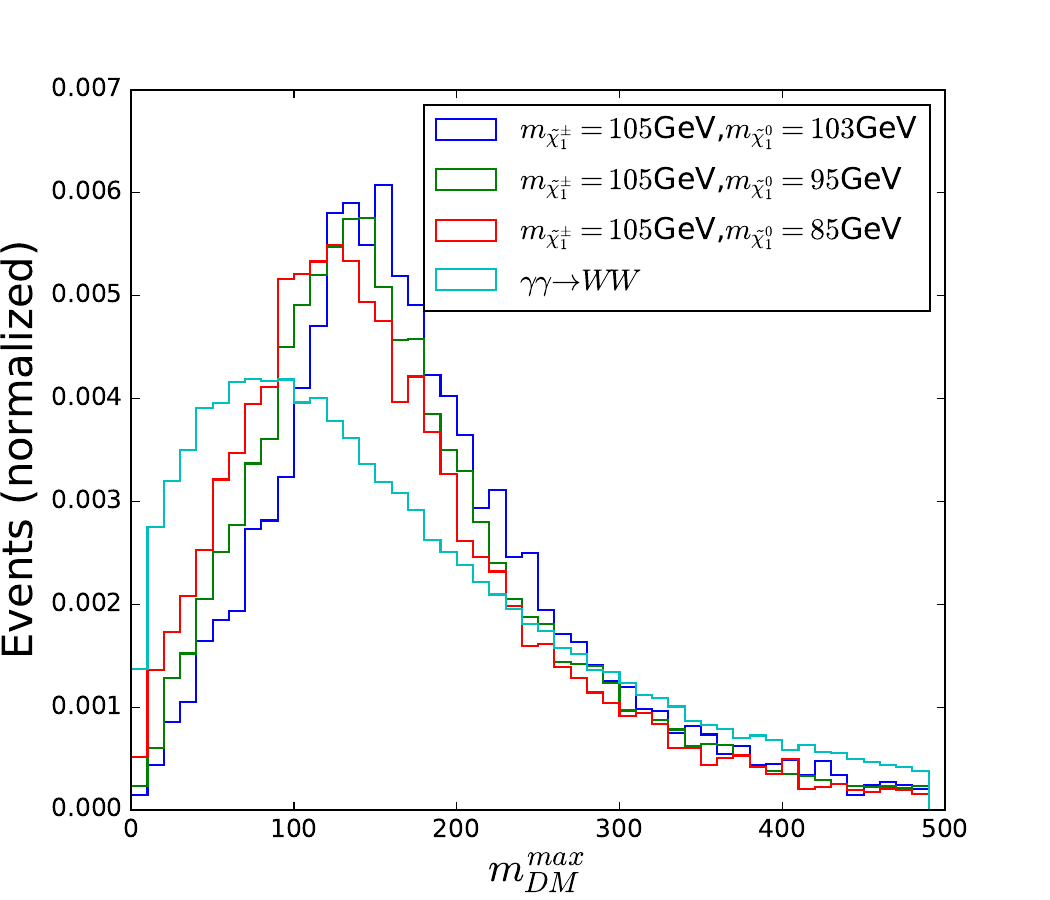}}
  \centerline{(c)}
\end{minipage}
\caption{Kinematical distributions of missing transverse energy $\slashed{E}_{T}$, transverse momentum of the leading lepton $p_{T}(\ell_{1})$ and the mass bound on the invisible system $m_{\text{DM}}^{\text{max}}$, for the SM background and the signal of three benchmark points: $m_{\tilde{\chi}^{\pm}_{1}}=105$ GeV with $\Delta m=m_{\tilde{\chi}^{\pm}_{1}}-m_{\tilde{\chi}^{0}_{1}}=2$, 10 and 20 GeV.}
\label{fig:dist}
\end{figure}

As an illustration, we present in \figurename~\ref{fig:dist} the distributions of the three kinematic variables for the signal Eq~\eqref{signal} of benchmark points $m_{\tilde{\chi}^{\pm}_{1}}=105$ GeV with $\Delta m=m_{\tilde{\chi}^{\pm}_{1}}-m_{\tilde{\chi}^{0}_{1}}=2$, 10 and 20 GeV, and for the SM background Eq~\eqref{bkg}. Compared with the invisible part in the background events consisting only of neutrinos, the distribution of $\slashed{E}_{T}$ for the signal events tends to peak around a higher energy region due to its massive component neutralinos, as shown in \figurename~\ref{fig:dist}\,(a). And as expected from this very argument, we can see from \figurename~\ref{fig:dist}\,(c) that the distributions for the benchmark points of larger $m_{\tilde{\chi}^{0}_{1}}(=85\,,95\,,103\,\text{GeV})$ peak around relatively larger mass values, which can also be well separated from the background. \figurename~\ref{fig:dist}\,(b) displays the distributions of the leading lepton $p_{T}$ from which we can see relatively quickly disappearing tails for the signal benchmarks, while the harder leptons from the $W$ decay in the background make for a long and less steep tail.

\begin{table}
\centering
\begin{tabular}{|l|c|c|}
\hline
  & \quad\makecell{$\Delta m$ = 2 GeV,\quad\quad \\ $m_{\tilde{\chi}^{\pm}_{1}}=105$ GeV\quad\quad} & \quad $W^{+}W^{-}$ events\quad\quad \\ \hline
No cuts applied & $6.52\times10^{-4}$ pb & $4.42\times10^{-3}$ pb \\ \hline
Diphoton \& Dilepton & $4.92\times10^{-5}$ pb & $6.06\times10^{-4}$ pb \\ \hline
Cut-1: $\slashed{E}_{T}\notin$\,[40,220] GeV & $3.01\times10^{-5}$ pb & $1.79\times10^{-4}$ pb \\ \hline
Cut-2: $P_{T}(\ell_{1})\in[2,5]$ GeV & $1.16\times10^{-5}$ pb & $6.19\times10^{-8}$ pb \\ \hline
Cut-3: $m^{\text{max}}_{\text{DM}}\in[110,260]$ GeV & $6.82\times10^{-6}$ pb & $4.42\times10^{-9}$ pb \\ \hline
\hline
\end{tabular}
\caption{Cutflow of cross sections for signal with $\Delta m$=2 GeV, $m_{\tilde{\chi}^{\pm}_{1}}=105$ GeV and for the SM background $W^{+}W^{-}$ events from photon fusion. }
\label{tab:cutflow}
\end{table}

Based on the above analysis of the kinematic distributions and focusing on the parametric space ranging over $m_{\tilde{\chi}^{\pm}_{1}}\in[105,275]$ GeV with $\Delta m\in[1,20]$ GeV, we apply three cuts with different criteria according to two signal regions (SR) with respect to the magnitude of the mass difference. As shown in \tablename~\ref{tab:cuts-1}, in the region of $\Delta m\in[1,15]$ GeV, we require $\slashed{E}_{T}\notin$\,[40,220]\,GeV, $P_{T}(\ell_{1})\in$[2,5]\,GeV and $m^{\text{max}}_{\text{DM}}\in[110,260]$\,GeV, while for a larger mass difference $\Delta m=20$ GeV, the cuts are applied at relatively higher values due to a more massive invisible system and harder leptons. Note that all three cuts are applied to events that have been pre-selected, that is, going through the `Diphoton \& Dilepton' procedure so that they are tagged by two intact protons and at least two oppositely charged leptons. In \tablename~\ref{tab:cutflow}, we present as an example the cross section cutflow for both the benchmark point of $m_{\tilde{\chi}^{\pm}_{1}}=105$ GeV, $\Delta m=2$ GeV and the SM background. Each row lists the effective cross sections after the corresponding cut in the first column. In the `No cuts applied' row, we show the original cross sections for the signal and background processes as in Eq.\eqref{signal} and \eqref{bkg}. `Diphoton \& Dilepton' row lists the effective cross sections after the pre-selection and the `Cut-1, 2, 3' correspond to the cuts listed in \tablename~\ref{tab:cuts-1} for the specific signal regions. As shown in the cutflow, after these cuts the SM background can be reduced to a negligible level compared to the signal, especially with the most efficient `Cut-2' for $P_{T}(\ell_{1})$.

\begin{table}
\centering
\begin{tabular}{|*{10}{c|}}
\hline
\diagbox{$\Delta m$}{$\alpha$}{$m_{\tilde{\chi}^{\pm}}$} & 105 & 115 & 125 & 135 & 145 & 155 & 165 & 175 & 185 \\ \hline
1 & 11.70 & 10.84 & 9.634 & 8.083 & 6.823 & 5.457 & 4.547 & 3.758 & 3.145 \\ \hline
2 & 32.48 & 31.99 & 27.26 & 24.32 & 20.60 & 17.55 & 15.19 & 12.43 & 10.79 \\ \hline
5 & 25.82 & 23.43 & 20.73 & 18.61 & 16.47 & 13.98 & 12.31 & 10.29 & 9.083 \\ \hline
10 & 9.903 & 9.072 & 8.117 & 6.878 & 6.136 & 5.499 & 4.806 & 3.899 & 3.297 \\ \hline
15 & 3.882 & 3.669 & 3.140 & 2.705 & 2.329 & 2.093 & 1.913 & 1.623 & 1.258 \\ \hline
20 & 1.572 & 1.456 & 1.283 & 1.170 & 1.045 & 0.900 & 0.775 & 0.676 & 0.557 \\ \hline
\end{tabular}
\begin{tabular}{|*{10}{c|}}
\hline
\diagbox{$\Delta m$}{$\alpha$}{$m_{\tilde{\chi}^{\pm}}$} & 195 & 205 & 215 & 225 & 235 & 245 & 255 & 265 & 275 \\ \hline
1 & 2.286 & 1.802 & 1.565 & 1.349 & 0.988 & 0.808 & 0.601 & 0.525 & 0.415 \\ \hline
2 & 8.910 & 7.485 & 5.830 & 4.911 & 4.012 & 3.311 & 2.576 & 2.133 & 1.716 \\ \hline
5 & 7.345 & 6.081 & 4.977 & 3.986 & 3.031 & 1.989 & 1.004 & $\backslash$ & $\backslash$ \\ \hline
10 & 2.674 & 2.265 & 1.882 & $\backslash$ & $\backslash$ & $\backslash$ & $\backslash$ & $\backslash$ & $\backslash$ \\ \hline
15 & 1.087 & 0.861 & $\backslash$ & $\backslash$ & $\backslash$ & $\backslash$ & $\backslash$ & $\backslash$ & $\backslash$ \\ \hline
20 & 0.500 & 0.381 & $\backslash$ & $\backslash$ & $\backslash$ & $\backslash$ & $\backslash$ & $\backslash$ & $\backslash$ \\ \hline
\hline
\end{tabular}
\caption{Significance achieved for integrated luminosity of 100 fb$^{-1}$ with identification of initial two photons and cuts on $\slashed{E}_{T}$, $P_{T}(\ell_{1})$ and $m_{\text{DM}}$ in \tablename~\ref{tab:cuts-1}. Systematic uncertainty is assumed to be 20\%. Units of $\Delta m$ and $m_{\tilde{\chi}^{\pm}}$ are GeV.}
\label{tab:sgnf-1}
\end{table}

\begin{figure}[t]
\centering
\includegraphics[scale=1.0]{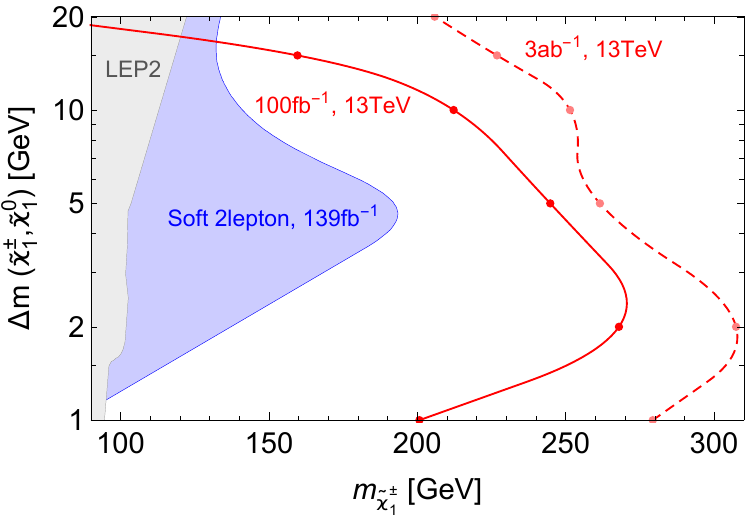}
\caption{$2\sigma$ exclusion limits based on event selection in \tablename~\ref{tab:cuts-1} for the signal process $pp\to p\left(\gamma\gamma\to\tilde{\chi}^{+}_{1}\tilde{\chi}^{-}_{1}\to\tilde{\chi}^{0}_{1}\tilde{\chi}^{0}_{1}W^{+}W^{-}\to\tilde{\chi}^{0}_{1}\tilde{\chi}^{0}_{1}\ell^{+}\nu_{\ell}\ell^{-}\bar{\nu}_{\ell}\right)p$ at 13 TeV LHC with the integrated luminosity of 100\,fb$^{-1}$\,(red solid line) and $3\,\text{ab}^{-1}$\,(red dashed line). Current experimental results are also presented including the ATLAS searches with soft lepton scenarios (blue region) \cite{Aaboud2018} and those from the LEP2 (grey region) \cite{Heister2002}.}
\label{fig:sgnf}
\end{figure}

With these cuts applied, we can finally achieve a promising search for the pair production of charginos via photon fusion at the 13 TeV LHC (Eq. \eqref{signal}). The results of the statistical significance at various parametric points are presented in \tablename~\ref{tab:sgnf-1}, which are calculated using the formula
\begin{align}
\alpha=S/\sqrt{B+(\beta B)^{2}},
\end{align}
with an integrated luminosity of 100\,fb$^{-1}$. $S$ ($B$) refers to the signal (background) events number after the event selection, while the systematic uncertainty $\beta$ is assumed to be 20\%. We can read from \tablename~\ref{tab:sgnf-1} that $2\sigma$ exclusion limits on $m_{\tilde{\chi}^{\pm}_{1}}$ can reach around 200, 270, 245, 210 and 160 GeV for $\Delta m=$ 1, 2, 5, 10 and 15 GeV, respectively. With more accumulated data of $\mathcal{L}=3\,\text{ab}^{-1}$, the $2\sigma$ exclusion limits on $m_{\tilde{\chi}^{\pm}_{1}}$ can be pushed to around 300\,GeV with $\Delta m\sim2$ GeV. \figurename~\ref{fig:sgnf} displays a fitted curve according the results in \tablename~\ref{tab:sgnf-1} and we also present the current limits from experiments including the ATLAS searches with soft lepton scenarios (blue region) \cite{Aaboud2018} and those from LEP2 (grey region) \cite{Heister2002}. It can be seen from \figurename~\ref{fig:sgnf} that the compressed region can be probed with good sensitivity. The $2\sigma$ exclusion bound on the mass of the lightest chargino $\tilde{\chi}^{\pm}_{1}$ varies from $200\sim270$ GeV in the range of mass splitting $1\,\text{GeV}<\Delta m(\tilde{\chi}^{\pm}_{1},\tilde{\chi}^{0}_{1})<10$ GeV at the 13 TeV LHC with the integrated luminosity of 100 fb$^{-1}$, while with a higher luminosity of 3 ab$^{-1}$ the exclusion limit can be extended to $250\sim308$ GeV in the same region of $\Delta m(\tilde{\chi}^{\pm}_{1},\tilde{\chi}^{0}_{1})$. The best probed regions are in the interval of $\Delta m(\tilde{\chi}^{\pm}_{1},\tilde{\chi}^{0}_{1})\in[2.0,2.5]$ GeV, while up to a relatively larger mass splitting of 20 GeV, previously unexplored space by ATLAS and LEP2 can also be reached ranging from $m_{\tilde{\chi}^{\pm}_{1}}\sim100$ GeV to 300 GeV.

As an interesting comparison, new angular cuts have been proposed recently for the higgsino search \cite{Baer:2021srt}, improving the 95\% CL exclusion limit for $m_{\tilde{\chi}^{0}_{2}}$ to $\sim325$ GeV for a mass splitting $\sim16$ GeV at the HL-LHC (3 ab$^{-1}$ and 14 TeV) which can cover the region beyond our results, while their sensitivity drops to $\sim200$ GeV for $\Delta m$ of a few GeV, which can, in turn, be complemented by our exclusion limits. The sensitivities to the Higgsino-like electroweakinos through soft-lepton events search have also been studied at the HE- and HL-LHC \cite{Canepa:2020ntc}. With a mass splitting in the range from about 3 to 50 GeV, the $2\sigma$ exclusion bound for $m_{\tilde{\chi}^{\pm}_{1}}$ ranges from 300 to 350 GeV, and this bound can be further improved by 60\% at the 27 TeV HE-LHC with the integrated luminosity of 15 ab$^{-1}$. While for a smaller mass splitting ($1\sim3$ GeV), our results (probing $m_{\tilde{\chi}^{\pm}_{1}}$ up to $\sim300$ GeV) can surpass the soft-lepton search at the HL-LHC, which can reach around 200 GeV. Future colliders such as ILC, CLIC and FCC-hh can improve the limits to the LSP mass by hundreds of GeV and even to TeV scale (for a review, see for example \cite{Canepa:2020ntc}).

\section{Conclusion}
In this paper, we demonstrate a search strategy for chargino pair production from photon fusion $pp\to p(\gamma\gamma\to\tilde{\chi}^{+}_{1}\tilde{\chi}^{-}_{1})p$ through the ultraperipheral collision of protons at the 13 TeV LHC, which is feasible by the aid of forward detectors tagging and measuring the outgoing intact protons near the colliding beams, including the AFP and CT-PPS located at the ATLAS and CMS, respectively. Measurement of the protons from UPC enables reconstruction of the initial information of the photon fusion and especially we adopt the mass bound on the invisible system, as well as transverse momentum of the leading lepton and missing transverse energy to explore the probe sensitivity in the compressed spectrum with nearly degenerate chargino and neutralino. $2\sigma$ exclusion limits can reach the range of $m_{\tilde{\chi}^{\pm}_{1}}\sim150$ GeV to 308 GeV for the mass difference $\Delta m(\tilde{\chi}^{\pm}_{1},\tilde{\chi}^{0}_{1})\in[1,15]$ GeV, with the largest probed $m_{\tilde{\chi}^{\pm}_{1}}$ at 270 GeV (308 GeV) for an integrated luminosity of 100 fb$^{-1}$ (3 ab$^{-1}$).

\acknowledgments
We would like to thank Yang Zhang for the helpful discussion on coding and using the PYLHE package. This work is supported by the Natural Science Foundation of the Higher Education Institutions of Jiangsu Province under grant No.\,22KJB140007.


\end{document}